\documentclass[sigconf, letter, 9pt]{acmart}

\usepackage{tikz}
\usepackage{amsmath}
\usepackage{xspace}
\usepackage{listings}
\usepackage{float}
\usepackage{multirow}
\usepackage{paralist}
\usepackage{todonotes}
\usepackage{caption}
\usepackage{comment}
\usepackage{cleveref}

\newcommand{\sys}{B-Side\xspace}

\newcommand*\BC[1]{\tikz[baseline=(char.base)]{
	\node[shape=circle,draw,inner sep=0.15pt] (char) {\textcolor{black}{#1}};}}

\newfloat{lstfloat}{htbp}{lop}
\floatname{lstfloat}{Listing}

\colorlet{punct}{red!60!black}
\definecolor{delim}{RGB}{20,105,176}
\colorlet{numb}{magenta!60!black}

\definecolor{background}{rgb}{0.97,0.97,0.97}
\definecolor{mGreen}{rgb}{0,0.6,0}
\definecolor{mGray}{rgb}{0.5,0.5,0.5}
\definecolor{mPurple}{rgb}{0.58,0,0.82}

\lstdefinelanguage{json}{
    basicstyle=\scriptsize\ttfamily\bfseries,
    keywordstyle=\color{magenta},
    numbers=left,
    numberstyle=\color{mGray},
    stepnumber=1,
    numbersep=5pt,
    showstringspaces=false,
    breaklines=true,
    breakatwhitespace=false,
    frame=lines,
    comment=[l]{//},
    morecomment=[s]{/*}{*/},
    commentstyle=\color{mGreen},
    backgroundcolor=\color{background},
    literate=
     *{0}{{{\color{numb}0}}}{1}
      {1}{{{\color{numb}1}}}{1}
      {2}{{{\color{numb}2}}}{1}
      {3}{{{\color{numb}3}}}{1}
      {4}{{{\color{numb}4}}}{1}
      {5}{{{\color{numb}5}}}{1}
      {6}{{{\color{numb}6}}}{1}
      {7}{{{\color{numb}7}}}{1}
      {8}{{{\color{numb}8}}}{1}
      {9}{{{\color{numb}9}}}{1}
      {:}{{{\color{punct}{:}}}}{1}
      {,}{{{\color{punct}{,}}}}{1}
      {\{}{{{\color{delim}{\{}}}}{1}
      {\}}{{{\color{delim}{\}}}}}{1}
      {[}{{{\color{delim}{[}}}}{1}
      {]}{{{\color{delim}{]}}}}{1},
}

\lstdefinestyle{cstyle}{
  language=C,
  basicstyle=\ttfamily\footnotesize, 
  keywordstyle=\color{blue},         
  commentstyle=\color{green},        
  stringstyle=\color{red},           
  numberstyle=\tiny\color{gray},     
  stepnumber=1,                      
  numbersep=5pt,                     
  showspaces=false,                  
  showstringspaces=false,            
  tabsize=4,                         
  captionpos=b,                      
  breaklines=true,                   
  breakatwhitespace=true,            
  frame=single,                      
  frameround=tttt,                   
  morekeywords={uint32_t, uint16_t, uint8_t} 
}

\ccsdesc[500]{Software and its engineering~Automated static analysis}
\ccsdesc[500]{Security and privacy~Operating systems security}

\keywords{Static analysis, Binary analysis, System call filtering}

\begin{document}

\setcopyright{none}
\acmYear{2024}\copyrightyear{2024}
\acmConference[MIDDLEWARE '24]{24th International Middleware Conference}{December 2--6, 2024}{Hong Kong, Hong Kong}
\acmBooktitle{24th International Middleware Conference (MIDDLEWARE '24), December 2--6, 2024, Hong Kong, Hong Kong}
\acmDOI{10.1145/3652892.3700761}
\acmISBN{979-8-4007-0623-3/24/12}

\title{\sys: Binary-Level Static System Call Identification}

\author{Gaspard Thévenon}
\affiliation{
    \institution{ENS Lyon}
    \city{Lyon}
    \country{France}
}

\author{Kevin Nguetchouang}
\affiliation{
    \institution{Grenoble INP}
    \city{Grenoble}
    \country{France}
}

\author{Kahina Lazri}
\affiliation{
    \institution{Orange Innovation}
    \city{Paris}
    \country{France}
}

\author{Alain Tchana}
\affiliation{
    \institution{Grenoble INP}
    \city{Grenoble}
    \country{France}
}

\author{\hphantom | Pierre Olivier}
\affiliation{
    \institution{The University of Manchester}
    \city{Manchester}
    \country{United Kingdom}
}

\begin{abstract}
System call filtering is widely used to secure programs in multi-tenant environments, and to sandbox applications in modern desktop software deployment and package management systems.
Filtering rules are hard to write and maintain manually, hence generating them automatically is essential.
To that aim, analysis tools able to identify every system call that can legitimately be invoked by a program are needed.
Existing static analysis works lack precision because of a high number of false positives, and/or assume the availability of program/libraries source code -- something unrealistic in many scenarios such as cloud production environments.

We present \sys, a static binary analysis tool able to identify a superset of the system calls that an x86-64 static/dynamic executable may invoke at runtime.
\sys assumes no access to program/libraries sources, and shows a good degree of precision by leveraging symbolic execution, combined with a heuristic to detect system call wrappers, which represent an important source of precision loss in existing works.
\sys also allows to statically detect phases of execution in a program in which different filtering rules can be applied.
We validate \sys and demonstrate its higher precision compared to state-of-the-art works: over a set of popular applications, \sys's average $F_1$ score is 0.81, vs. 0.31 and 0.53 for competitors.
Over 557 static and dynamically-compiled binaries taken from the Debian repositories, \sys identifies an average of 43 system calls, vs. 271 and 95 for two state-of-the art competitors.
We further evaluate the strictness of the phase-based filtering policies that can be obtained with \sys.
\end{abstract}

\maketitle

\section{Introduction}

System call filtering~\cite{SECCOMP} is widely used to sandbox programs in various environments such as containers~\cite{GVISOR_SECCOMP}, desktops~\cite{FLATPAK_FILTER, FIREJAIL_FILTER} and mobile~\cite{ANDROID_SECCOMP} systems.
The current methods to define filtering rules are suboptimal.
On the one hand, creating generic rules that work for most applications trades off security for generality and leaves a large attack surface: for example, Android's filtering rule applies to all applications but only blocks 17 of the 271 arm64 Linux system calls~\cite{ANDROID_SECCOMP}, and Docker blocks by default only 43 of the 350+ x86-64 system calls~\cite{CONFINE}.
Similarly, modern desktop software deployment and package management systems relying on system call filtering to sandbox applications fail to provide strict generic policies: Flatpak's filtering rule allows completely unrestricted access to 308 system calls for x86-64~\cite{FLATPAK_FILTER}, and the Firejail~\cite{FIREJAIL} sandbox used by AppImage only precludes 15 system calls by default~\cite{FIREJAIL_FILTER}.
Service (e.g. cloud) providers may create system call filtering policies on the basis of their needs and present them as a take-it-or-leave-it basis to clients, but these policies are either too generic to bring strong safety guarantees, or too strict to be applicable to a broad set of applications.
On the other hand, manually creating filtering rules that are tailored in a per-application fashion~\cite{QEMU_SECCOMP, CHROMIUM_SECCOMP, OPENSSH_SECCOMP, FIREFOX_SECCOMP, JENNY} is highly secure but also very difficult to achieve and hard to maintain~\cite{ORACLE_SECCOMP, VSFTPD_BUG}, as it requires detailed knowledge about the application and can only be achieved by an expert such as the application's developer.

The solution to create tailored rules without manual effort is to automate that process: there is a need for tools that can identify the list of all system calls that can be made by an application.
This task is made hard by the fact that in many environments (e.g. cloud), user applications are only available in a binary form and a cloud provider or desktop sandbox wishing to enforce a system call filtering policy cannot assume access to the sources of applications or libraries~\cite{ABHAYA}.
Dynamic analysis~\cite{DYNAMIC1, DYNAMIC2, DYNAMIC3, DYNAMIC4, UNIKRAFT_LOGIN} is a poor fit to build per-application system call allowed/denied lists, due to the unavoidable risk of false negatives stemming from the difficulty to achieve full coverage.
Existing tools based on static analysis either do not achieve good coverage~\cite{TSAI, HERMITUX}, lack flexibility and generality~\cite{HERMITUX, CHESTNUT, CONFINE, SYSFILTER, ENTER_SANDBOX, WAGNER_STATIC, SAPHIRE, ABHAYA, BASTION} by requiring access to the application/library sources and/or being restricted to a subset of applications written in specific languages, or suffer from lack of precision due to suboptimal value tracking strategies during the system call identification process~\cite{CHESTNUT, SYSFILTER}.

We propose \sys (for Binary-level System call Identifier), \emph{a static binary analysis framework that aims to automatically and precisely identify all the possible system calls that can be issued by a given x86-64 statically or dynamically compiled ELF executable.}
\sys targets compiled executables: it makes no assumption on the availability of application/library sources and works independently of the language the application is written in.
To that aim, \sys disassembles the target binary and uses symbolic execution, a technique which was proven effective at reducing attack surface~\cite{LIGHTBLUE}, to determine the types of system calls that can possibly be invoked at each of the program's system call sites.

\sys tackles several challenges, the main one being comprehensive and precise system call identification.
Being comprehensive means that every system call that may be invoked by the analyzed binary should be identified: any false negative (system call missed by the analysis but legitimately called at runtime) will result in a legitimate system call being flagged at runtime as illegal by a filtering rule derived from the analysis -- something unacceptable in production.
Being precise signifies that the set of system calls identified should be as small as possible, as false positives (system calls identified by the analysis but never invoked at runtime) limit the strictness of filtering rules derived from the analysis.
\sys tackles these challenges by using a search algorithm navigating the program's CFG backward from each system call site, and leveraging symbolic execution to determine every possible value that can be present in the ABI-defined CPU register containing the system call identifier at the time of invocation.
Further, \sys uses a heuristic to detect occurrences of system call wrappers and drives symbolic execution appropriately in such cases, avoiding the significant loss of precision of existing works~\cite{CHESTNUT, SYSFILTER} in the presence of such wrappers, which are widely used in standard libraries and runtimes.

Recent works have presented advanced system call filtering policies: temporal system call specialization~\cite{TEMPORAL} installs different filters for different phases of execution of an application, and system call flow integrity enforces the ordering of certain system calls~\cite{SYSCALL_FLOW, STATEFUL_SECCOMP} or a valid control flow in the program prior to a system call invocation~\cite{BASTION}.
We explore \sys's capacity to analyze binary-only programs for generating such advanced filtering policies.
This is achieved through the construction of an automaton in which states represent phases of execution of the program, and transitions between states correspond to invocation of certain system calls.
For each phase \sys identifies the list of possible system calls that may be invoked, as well as the system calls triggering transitions to other phases.

We validate \sys by showing the absence of false negatives, and evaluate its precision (amount of false positives) against two competitors over 557 Debian 10 x86-64 ELF binaries.
Over more than 200 dynamically-compiled binaries, \sys identifies an average of 55 system calls, vs. 274 for Chestnut and 96 for SysFilter.
We further evaluate the precision of the phase detection feature of our tool, demonstrating over an example that a phase-based filtering rule derived from the automaton would lead a reduction of 11 to 13\% of the system call types allowed compared to a vanilla filtering rule enforced over the entire program's lifetime.

This paper makes the following contributions:
\begin{compactenum}
\item We list and detail the various challenges arising in the domain of system call identification.
\item We design, implement, validate and evaluate \sys, a static binary analysis framework aiming at automatically identifying all the possible system calls that can be issued by a given x86-64 statically or dynamically compiled ELF executable, as well as phases of execution for such executables during which different sets of system calls may be invoked.
\end{compactenum}

\section{System Call Identification: Background and Challenges}
\label{sec:challenges}

\subsection{Dynamic vs. Static Analysis}
Dynamic analysis can be used to address the problem of system call identification~\cite{FACECHANGE, KURMUS_NDSS, MINING_SANDBOXES, CIMPLIFIER, CONT_DEBLOAT, DOCKER_SLIM, PRACTICAL}, however this class of techniques is well-known to struggle to achieve complete coverage~\cite{CONFINE}, and thus suffer from false negatives: these approaches under-estimate the set of system call types that can be issued by an application at runtime.
This leads to normal application behavior being classified as malicious which is unacceptable in most situations in production.
As a result, several works rather rely on static analysis~\cite{CONFINE, SYSFILTER, CHESTNUT, ENTER_SANDBOX, WAGNER_STATIC} which, when implemented properly, avoids the issue of false negatives.
Still, static analysis overestimates and identifies a superset of the system calls that will be invoked at runtime by a program, it suffers from false positives.
This reduces the strictness of the filtering policies that can be derived, but does not represent a blocking issue like false negatives.
For that reason, \sys relies on static analysis.

Static analysis tools work on the application binary or sources~\cite{CHESTNUT, ENTER_SANDBOX, SYSFILTER, CONFINE, WAGNER_STATIC, BASTION}.
Most make the assumption that program and/or library sources are available~\cite{WAGNER_STATIC, CONFINE, ABHAYA, SAPHIRE, BASTION, C2C}.
We argue that access to sources is an unrealistic assumption in many situations.
For example, a cloud provider executing its clients' applications in containers rarely has access to the source code of client programs and library dependencies.
Further, source-level analysis does not scale to large numbers of applications as it is generally language-specific and needs to be tailored toward all the source-level ways to invoke system calls:
language-specific wrappers, assembly code (potentially inline), etc.
As previously stated, \sys operates only on binaries and does not assume access to the sources of applications or libraries.

In essence, static binary analysis consists in:
1) disassembling the target binary;
2) recovering the program's Control Flow Graph (CFG);
3) identifying system call sites;
and 4) for each site inferring what possible system calls can be issued at that point.
Several of these steps involve addressing non-trivial challenges that are discussed below.

\subsection{Disassembling}
Disassembling arbitrary binaries is a hard problem~\cite{RAMBLR, SYSFILTER}, due in particular to the difficulty to differentiate code from data in some scenarios~\cite{CODE_DATA_ASM}, and to determine function boundaries~\cite{FUNC_BOUNDARIES}.
Still, popular modern compilers such as GCC/LLVM do not mix code and data and embed by default metadata in binaries (including stripped ones), such as stack unwinding information, helping in recovering function boundaries, and, as a result, modern binary analysis tools can tackle these issues for most executables.

\subsection{CFG Recovery}
\label{sec:challenge2}
The majority of the machine-code CFG of an executable can be recovered from disassembled code by observing direct control flow instructions taking as operand an immediate address in the code segment.
The difficulty lies in resolving indirect calls, i.e. calls and jumps to a memory operand that cannot easily be determined statically~\cite{SYSFILTER, ANGR}, resulting from the use of constructs such as function pointers in the sources.
As a result some edges may be missing in the disassembled CFG used by the system call identification tools, which in turn may lead to false negatives due to missed system calls.
This challenge is partially addressed through heuristics, that over-estimate the CFG by creating edges from each basic block ending in an indirect call toward each possible \emph{address taken}, i.e. each address in the code segment that is used as the source of a store operation~\cite{SYSFILTER} (function pointer assignments).
These can be complemented by other techniques including symbolic execution, backwards slicing, etc~\cite{ANGR}.

\subsection{Challenges: Comprehensive and Precise System Call Identification}
Once the binary is disassembled, identifying system call sites is straightforward: it consists in finding each occurrence of the \texttt{syscall} x86-64 instruction that is used to perform a system call.
Understanding \emph{what} system call(s) can be invoked at each site is more difficult and represents the key challenge tackled by \sys.

The Linux ABI~\cite{SYSTEMV_ABI} states that when the \texttt{syscall} instruction is invoked, the \texttt{\%rax} register should contain an integer identifying the system call to be executed.
Hence, the system call identification consists in determining the possible value(s) present in \texttt{\%rax} at the time each \texttt{syscall} instruction in the code segment is invoked.
Due to the nature of system call invocation, in most situations a constant value (such as an immediate) will be loaded in that register, we thus can hope to determine it statically.

\paragraph{Naive and Use-Define Chains-based Methods.}

\begin{figure}
    \center
    \includegraphics[width=0.45\textwidth, page=1]{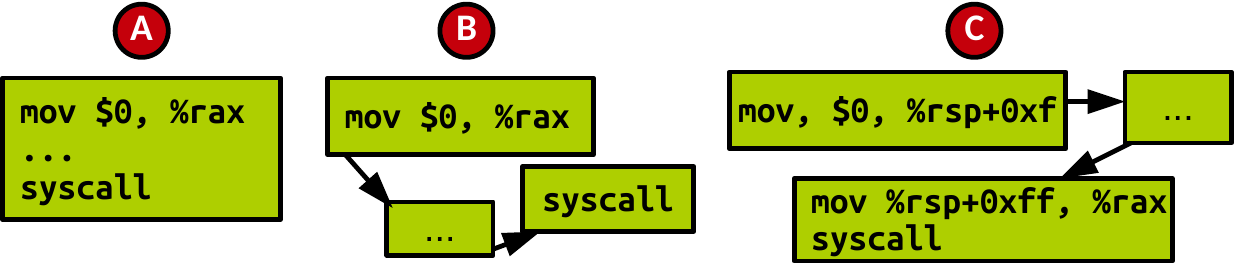}
    \caption{Scenarios in which the system call type defining immediate is: in the same basic block as the \texttt{syscall} instruction (A); in a different basic block (B); and with the immediate value propagated through memory on the stack (C).}

    \label{fig:graph1}
\end{figure}

A naive method such as the one used by Angr~\cite{ANGR} or Confine~\cite{CONFINE} does not take the CFG into account: for each occurrence of \texttt{syscall}, it considers only the containing basic block or direct predecessors and performs simple analysis to try to determine the value of \texttt{\%rax} at the time of \texttt{syscall} (Figure~\ref{fig:graph1} \BC{A}).
Such an approach is not sufficient to identify all the system calls made in most programs because the immediate defining a system call identifier may originate from a basic block different from the one containing the \texttt{syscall} instruction, or from a direct predecessor (Figure~\ref{fig:graph1} \BC{B}).

Another method considers the CFG and use-define chains~\cite{COMPILER_BOOK} to track the immediate that can be placed in \texttt{\%rax}~\cite{SYSFILTER}.
This approach is also limited as it fails to track the propagation of immediates through memory.
If an immediate is for example stored on the stack before being loaded into \texttt{\%rax} prior to \texttt{syscall} invocation (Figure~\ref{fig:graph1} \BC{C}), this method will fail to identify this value~\cite{SYSFILTER}.
This is problematic in many cases, for example with the Go language where system calls wrappers functions, passing arguments on the stack~\cite{GO_CONVENTION} are used.
As a result existing works relying on that technique require access to the sources of all the application/library code passing system call identifiers through memory~\cite{SYSFILTER}.

\paragraph{Symbolic Execution.}
To address these issues we propose in \sys to leverage symbolic execution~\cite{SYMBOLIC_EXECUTION}, navigating the CFG to determine the value of all the immediates that may end up in \texttt{\%rax} when \texttt{syscall} is invoked: these can be tracked even if they pass through memory.
Using symbolic execution to identify system calls have been hinted at in past work, although there is no in-depth study of this technique to tackle that particular problem: although Chestnut~\cite{CHESTNUT} mentions it, in practice it is only used for the disassembly and not the system call identification phase, for which the tool manually tracks operands of \texttt{mov} and \texttt{xor} instructions prior to \texttt{syscall}\footnote{\url{https://github.com/IAIK/Chestnut/blob/91269361e53993d7e9e5acb0614559774694ba35/Binalyzer/syscalls.py\#L45}}.

\begin{figure}
    \center
    \includegraphics[width=0.45\textwidth, page=2]{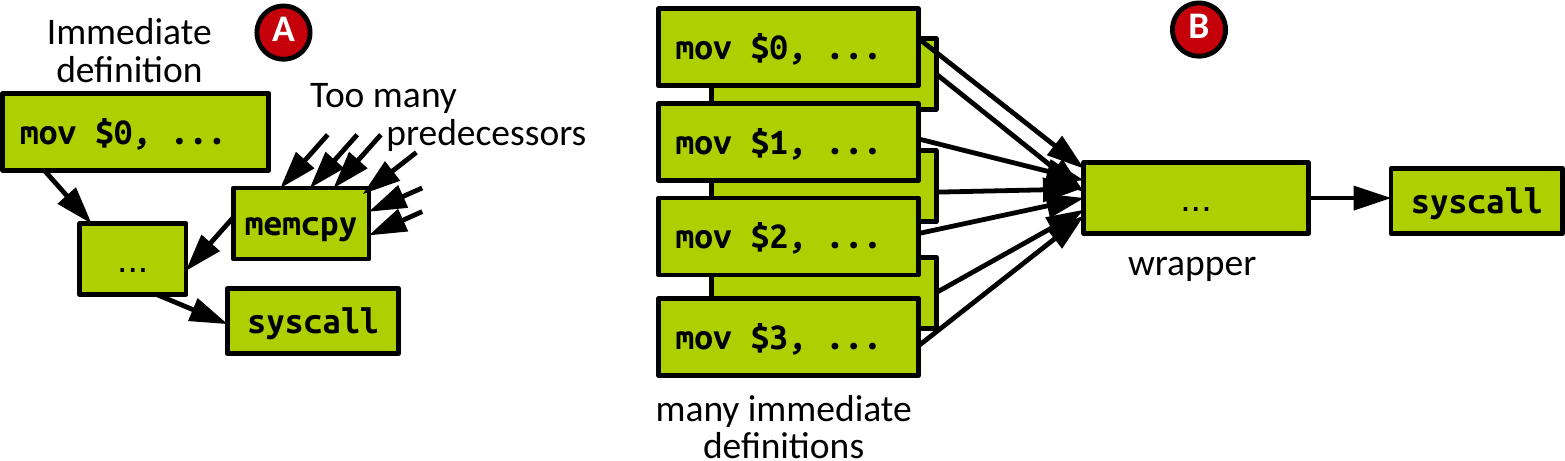}
    \caption{A function with many predecessors between the immediate definition and the \texttt{syscall} instruction make symbolic exploration difficult (A);
    a system call wrapper function leading to a high over-estimation of the system call set (B).}

    \label{fig:graph2}

\end{figure}

Still, symbolic execution comes with its own set of challenges, including combinatorial explosion: when the path explored is too large, the computational complexity and/or memory footprint makes that the execution will never finish or will crash.
This prevents an exhaustive forward symbolic exploration of the entire CFG from the program entry point.
Hence, we perform a search starting from system call sites and going backwards towards immediate definitions.

Even with backward exploration, other problems arise: when a function called from many places in the binary (e.g. \texttt{memcpy}) is called in between the immediate definition and the corresponding \texttt{syscall} instruction, the search may lead to a state explosion because of the numerous predecessors of the popular function call site. (Figure~\ref{fig:graph2} \BC{A}).
With \sys we use an advanced technique in which we search backwards in the CFG for nodes preceding the one containing \texttt{syscall}.
We start \emph{forward} symbolic execution from each predecessor, and direct the symbolic search toward the nodes identified as \texttt{syscall} predecessors during the backwards search.
This allows avoiding the unneeded and costly exploration of popular functions' predecessors.

Another important issue with symbolic execution is that of system call \emph{wrappers}, i.e. functions in the program's source code encapsulating system calls.
The wrapper generally takes the system call number as parameter, followed by the system call arguments list.
A good example is the \texttt{syscall()} function from the C standard library, but other examples can be found in languages such as Go and Rust, as well as in programs such as Valgrind and ptrace-based monitors.
They are generally called from multiple places in a program, hence they lead to the state explosion issue described above.
Further, because many (and potentially all) system call types can be made through the wrapper, a backward search from the \texttt{syscall} instruction within the wrapper up to the immediates definitions will identify a very large number of system calls: all those that can be called through the wrapper, independently of the ones actually made by the program (Figure~\ref{fig:graph2} \BC{B}).
In \sys we use a heuristic to detect such wrappers and start the backward search from the wrapper invocation call sites rather than from \texttt{syscall} instruction, 
avoiding the state explosion and the system call overestimation issues by limiting the predecessors explored.

\subsection{Soundness}
None of the existing binary system call identification studies can guarantee full soundness (the absence of false negatives): for example, the identification process implemented by SysFilter~\cite{SYSFILTER} is based on intra-procedural analysis, leading to false negatives as the tool does not support situations in which the function declaring the constant identifying a system call is different from the function containing that system call's invocation (system call wrappers). 
Chestnut~\cite{CHESTNUT} leverages Angr's symbolic execution engine to recover the application's CFG, a technique for which full correctness is not guaranteed~\cite{ANGR_CFG}.
Similarly, we do not claim full soundness with \sys.
We detail \sys's barriers to soundness in Section~\ref{sec:soundness}, along with arguments why it remains a useful tool in many scenarios and how it advances the state of the art on certain soundness-related challenges.

\section{Related Works}
\label{sec:related-works}

Many tools have relied upon dynamic analysis~\cite{DYNAMIC0, DYNAMIC1, DYNAMIC2, DYNAMIC3, DYNAMIC4, PRACTICAL, MINING_SANDBOXES}.
As previously mentioned, given the limitations of this class of techniques, we rather focus on static analysis~\cite{WAGNER_STATIC, TSAI, HERMITUX, CONFINE, SYSFILTER, CHESTNUT, ABHAYA, SAPHIRE}.
Some of these works can afford false negatives~\cite{HERMITUX, TSAI} because their goal is not system call filtering rule generation, hence they are unfit for our aims.
Some works also assume full access to the applications' and/or library dependencies' sources~\cite{WAGNER_STATIC, CONFINE, C2C}.
SysPart~\cite{SYSPART} is able to automatically identify phases of execution during which different filtering rules can be applied, however it can only do so on a very small set of applications: servers software with warmup vs. serving phases.

Here we develop on 2 systems that
1) leverage static analysis,
2) aim at avoiding false negatives for security reasons and
3) work on binaries only, i.e. do not assume access to either application or library dependencies' sources.

\paragraph{SysFilter.}
SysFilter~\cite{SYSFILTER} focuses on precise disassembly and CFG recovery.
Precise disassembly of the target binary is obtained by leveraging stack unwinding information, present even in stripped binaries.
An overestimation of a precise CFG can also be recovered, including indirect calls, by using the following heuristic: the binary is first scanned to search for \emph{addresses taken}, i.e. addresses in the code segment used as operand of a store operation (e.g. assignment of function pointers).
Indirect calls in the CFG are then resolved as being made to every possible address taken.
\sys takes inspiration from this technique to recover a precise model of the CFG.
SysFilter's system call identification method is quite simplistic, analyzing the machine code to determine \texttt{\%rax}'s value at the time of \texttt{syscall} with use-define chains in an intra-procedural fashion.
As a result the tool fails to determine system calls when the immediate defining a system call type travels through memory between functions, which is the case for example with system call wrappers, used in a non-negligible amount of binaries.
This leads to SysFilter suffering from false negatives.
Another limitation is that it only works on dynamically compiled binaries and does not support (non-PIC) static ones.
On the contrary, \sys uses inter-procedural symbolic execution for system call identification, which combined with our system call wrapper detection method leads to a much more robust approach that allows, as demonstrated in our evaluation, a better precision and higher compatibility with arbitrary binaries, dynamically as well as statically compiled.

\paragraph{Chestnut.}
Chestnut~\cite{CHESTNUT} relies on symbolic execution for disassembly only, and performs for system call identification a simple value tracking analysis on operands of \texttt{mov} and \texttt{xor} instructions (in registers only) prior to \texttt{syscall} in a backwards fashion.
As we demonstrate in our evaluation (Section~\ref{sec:evaluation}), this analysis is limited and fails in various scenarios.
First, the limited scope of the exploration (30 instructions) is not sufficient to tackle many executables and libraries.
Second, undirected backwards symbolic execution suffers from state explosion/lack of precision in the case of popular functions/system call wrappers.
Chestnut's binary analysis includes a hardcoded detector for the Glibc's wrapper, but we confirmed that it is not able to identify system calls in other languages (such as Go) or libcs (e.g. Musl) using different wrappers.
Chestnut hence relies on dynamic analysis to refine the binary analysis results, which defeats the purpose of using static analysis in the first place.
\sys uses only static analysis, is compatible with arbitrary binaries, with a heuristic able to identify system calls wrappers in arbitrary languages.
Furthermore, \sys actually relies on symbolic execution for system call identification, a technique much more robust and far-reaching compared to that of Chestnut.

\section{\sys: Binary-Level System Call Identification}
\label{sec:contribution}

\subsection{Design's Objectives and Assumptions}

We address the following research question:
relying solely on static binary analysis, can we build a system call identification tool that is 1) valid and 2) precise?
Validity implies the absence of false negatives, i.e. system calls missed by the analysis but invoked legitimately by the program at runtime.
Precision requires the minimization of false positives, i.e. system calls identified by the analysis, but that will never be invoked at runtime.

In that context we make the following assumptions.
We assume no access to sources.
The target programs can be written in arbitrary languages, built with arbitrary compilers, and use arbitrary standard libraries (e.g. libc, Go, Rust) possibly implementing various wrappers encapsulating system call invocations.
We do not rely on any kind of dynamic analysis whatsoever.
Regarding the disassembling process, we assume that the target binaries are not obfuscated.
Although there is probably a certain amount of proprietary software running obfuscated e.g. in the cloud, we presume that the majority of workloads do not include obfuscated binaries, as such techniques generally impact performance negatively.
We also assume that the disassembler used can correctly identify all the code as well as determine function boundaries.
We do not assume that the disassembler can resolve all indirect jumps.
We focus on compiled programs and scope out interpreters/managed runtime (e.g. Python, the JVM, etc.) which should be handled very differently.
We rely on symbolic execution, hence the analysis may be slow or even never terminate due to the well-known path explosion issue.
Although we employ optimizations to avoid these problems, we cannot eliminate them completely, and we assume it is acceptable for \sys to time out on a subset of programs.

\subsection{Overview}

\begin{figure}
    \center
    \includegraphics[width=0.35\textwidth]{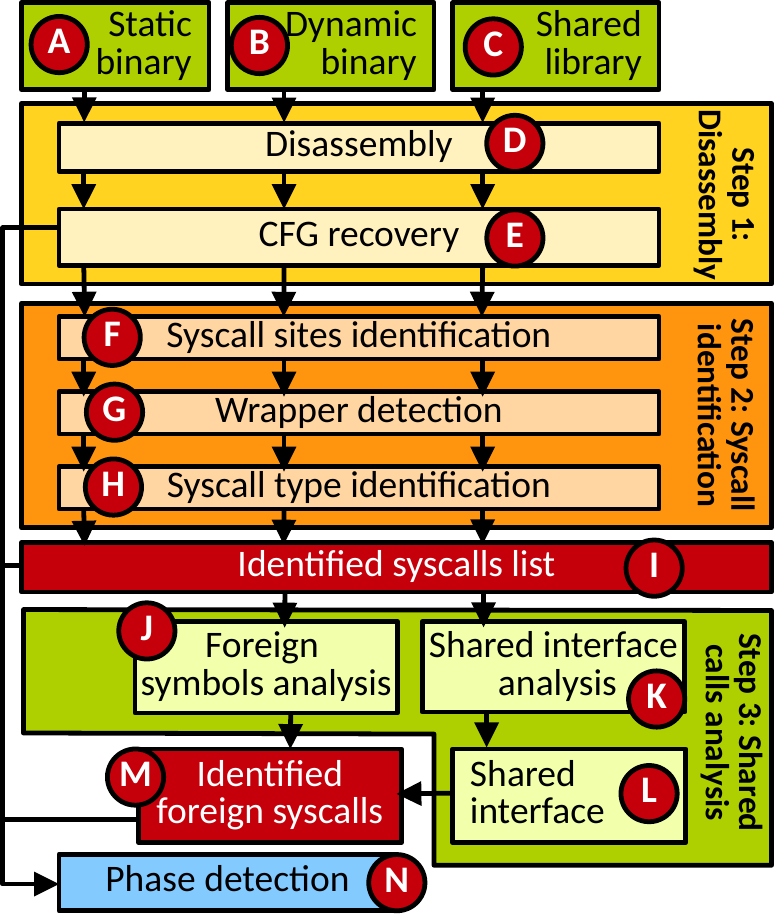}
    \caption{Overview of \sys's system call identification process divided into 3 main steps.}

    \label{fig:overview}
\end{figure}

Figure~\ref{fig:overview} presents an overview of the analysis flow.
The tool takes as input static~\BC{A} and dynamically compiled executables~\BC{B} with their shared library dependencies~\BC{C}.
In the first step of the process, binaries are disassembled.
A first basic disassembly phase~\BC{D} outputs a CFG of machine code basic blocks.
For the vast majority of programs/libraries this CFG is incomplete as indirect jumps are not fully resolved, hence we use a heuristic to conservatively create new edges in the CFG~\BC{E} to ensure that the next steps of the process will be able to explore all the possible paths in the program/library.
The second step of the process regards system call identification.
The system call sites are identified by scanning basic blocks for the \texttt{syscall} x86-64 instructions~\BC{F}.
Next, for each call site, we use a heuristic to determine if it corresponds to a system call wrapper~\BC{G}.
Finally, for each system call site, we proceed to identify what are the types of system calls that can be invoked at that site~\BC{H}.
To that aim we leverage symbolic execution and use a strategy that slightly differs according to the wrapper/non-wrapper status of the call site in question.
We then obtain the list of system calls that may be invoked by the program/library~\BC{I}.
For dynamically compiled programs, a last step consists in determining the system calls invoked as part of shared library calls.
To do so, we analyze the functions exposed by the library dependencies of the executable~\BC{K} and list, for each library, the possible system calls invoked by each exposed function.
This metadata is written in a file created for each library that we name the \emph{shared interface}~\BC{L}.
We then list the shared library calls made by the dynamically compiled program~\BC{J}, and, with the help of the shared interface, we can then output the list of system calls made by the program through shared libraries~\BC{M}.
Finally, the precise CFG is combined with the list of identified system calls at each site to detect phases in the program's execution~\BC{N}: at runtime, a different system call allowed list can be enforced during each phase.

\sys is implemented in Python and consists in 3217 lines of code.
In the rest of this section, we detail each step of the system call identification process.

\subsection{Step 1: Disassembly}

\paragraph{Disassembly.}
Taking the target binary/library as input, \sys starts by disassembling it (\BC{D} in Figure~\ref{fig:overview}) using Angr~\cite{ANGR}.
Angr uses the Capstone~\cite{CAPSTONE} disassembly engine which is quite robust~\cite{SOK_DISASSEMBLY} and allows in particular to separate code from data and to identify function boundaries, addressing the \emph{proper disassembly} challenge listed in Section~\ref{sec:challenges}.
The output of that step is the machine code composing the program/library, organized into basic blocks that are linked into a CFG.
A well-known limitation of modern disassembly tools is that they struggle with indirect call resolution, leading to missing edges in the CFG.
We address that issue in the next step of the disassembly phase: CFG recovery.

\paragraph{CFG Recovery.}
The next step is to recover the CFG (\BC{E} in Figure~\ref{fig:overview}).
Although Angr/Capstone makes some attempts at resolving indirect jumps targets~\cite{ANGR, SOK_DISASSEMBLY}, we observe that the result is not perfect.
Concretely, many edges corresponding to function pointers calls are missing in the CFG of most programs/libraries.
Many programs and libraries (e.g. the C standard library, used by the majority of executables) make extensive use of function pointers, hence it is important for \sys to tackle this challenge if we wish for it to be applicable beyond a small set of restricted programs/libraries that do not use function pointers.

We thus rely on a heuristic to resolve indirect calls.
We take inspiration from SysFilter~\cite{SYSFILTER}, which defines the concept of \texttt{address taken}.
An address taken is an absolute address identified in the disassembled code as located within the code segment at runtime, which is used as the operand of a load operation.
This indicates the assignment of a function pointer.
In x86-64 this will concretely appear in the form of a Load Effective Address (\texttt{lea <address taken>}) instruction.
The key idea is to resolve the possible targets of each indirect jump to be the list of all addresses taken in the binary.
This represents a conservative overestimation of the locations that can be reached from a basic block ending with an indirect jump instruction.

\begin{figure}
    \center
    \includegraphics[width=0.45\textwidth]{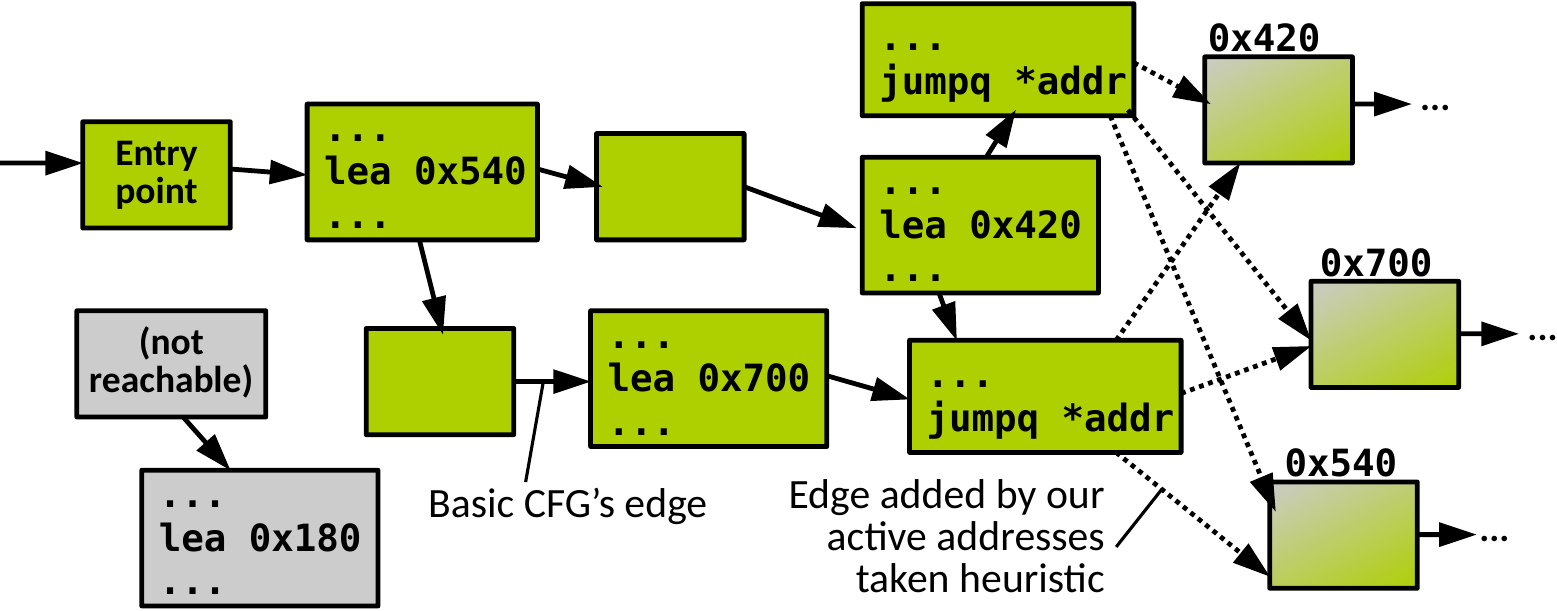}
    \caption{We use active addresses taken, i.e. addresses operands of \texttt{lea} instructions reachable from the program's entry point, to overestimate the list of indirect jump targets.}

    \label{fig:active-addresses-taken}
\end{figure}

Although the number of addresses taken is generally relatively low (a few tens of addresses in most binaries), we observe that it still gives a relatively large overestimation of the CFG, translating into an overestimation of the system calls identified.
To tackle that issue, we refine the notion of address taken and introduce the concept of \emph{active addresses taken}.
The set of active addresses taken in a program/library represents the subset of addresses taken that are loaded with \texttt{lea} in CFG nodes that are reachable from the program's entry point or from the library's exposed functions invoked by a given program.
Identifying active addresses taken is an iterative process.
For a program, we start with the basic CFG obtained from Angr/Capstone.
We then identify a first set of active addresses taken starting from the entry point.
Next we update the CFG's indirect calls with this first set of addresses taken, and restart the process from the entry point as new active addresses taken may be discoverable given the new edges added to the CFG.
This process, illustrated in Figure~\ref{fig:active-addresses-taken}, continues until no new active addresses taken are found and the precise CFG is obtained.
For a library, we first identify the functions exposed by a library that are invoked by a given dynamically compiled executable.
To that aim we list the library's functions that are reachable from the program's entry point once we have obtained its precise CFG.
We then repeat the process of finding active addresses taken within the library, by considering as entry points each of the exposed functions reachable by the considered program.
This tackles the aforementioned challenge regarding precise CFG recovery.

\subsection{Step 2: System Call Sites and Type Identification}

After disassembling and constructing a precise CFG, the next step consists in identifying all the system call sites locations (\BC{F} in Figure~\ref{fig:overview}).
Assuming a precise disassembly, this is a straightforward operation, consisting in finding all occurrences of the \texttt{syscall} instruction.
At that point we assume that the CFG obtained from the previous step is correct and consider only the occurrences that are reachable from the program's entry point (or exposed function for shared libraries).

\begin{figure}
    \center
    \includegraphics[width=0.45\textwidth]{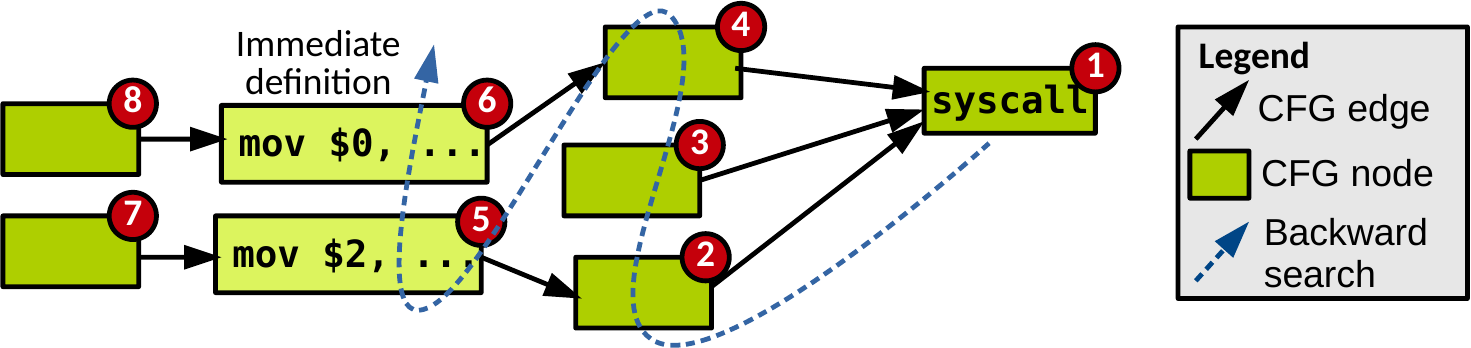}
    \caption{System call identification process: starting from a call site, predecessors are selected in BFS mode one after the other as the start node of a forward symbolic execution search. This process continues until all reachable nodes for which the symbolic execution is able to determine the system call type have been found.}
    \label{fig:search}
\end{figure}

The next step is to identify, for each considered \texttt{syscall} occurrence, the possible system call type(s) that can be made at that particular point.
That process (\BC{H} in Figure~\ref{fig:overview}) is slightly different depending on the \texttt{syscall} occurrence being part of a system call wrapper or not (\BC{G}).
Here we describe how that is achieved when the call site is not part of a wrapper.
Wrappers will be discussed next.

Identifying the system call type(s) that can be invoked by a \texttt{syscall} occurrence that is not a wrapper is illustrated in Figure~\ref{fig:search}.
For each system call site (e.g. \BC{1} in Figure~\ref{fig:search}), \sys selects predecessor nodes in a Breadth-First Search (BFS) manner, in our example starting with \BC{2}.
For each of these predecessors $p$, \sys starts a forward symbolic execution process from $p$ to the system call site.
In order to speed up the symbolic execution process and avoid the search getting lost in paths not leading to the system call site, we use directed symbolic execution and aim to reach the nodes leading to the \texttt{syscall} instruction, i.e. the nodes we have already identified in the BFS backwards search.

Once the system call site is reached by the symbolic search, we query the value in \texttt{\%rax}.
If that value is \emph{symbolic}, as will be the case when we start from \BC{2}, \BC{3} or \BC{4}, the symbolic search was not able to determine a concrete value for \texttt{\%rax}'s content.
In such case \sys continues with the backwards BFS process.
The symbolic search will succeed once it starts from a node which defines an immediate that will end up in \texttt{\%rax} at the call site.
It is the case for nodes \BC{5} and \BC{6} in our example in Figure~\ref{fig:search}.
The backwards search process stops for a given path once such an immediate-defining node has been found.
For example once the symbolic search starting from \BC{6} is successful, the backwards search does not explore \BC{8}.
Similarly, we do not explore \BC{7} because \BC{5} is immediate-defining.
Once all such immediate-defining nodes directly reachable from the call site have been found, we can guarantee that we determined the value of \texttt{\%rax} for every execution path reaching the call site.
We output the list of all possible system calls that can be made at that site: in our example that would be 0 (\texttt{read}) and 2 (\texttt{open}).

\paragraph{System Call Wrapper Detection Heuristic.}

As mentioned above, detecting system call wrappers is important for a comprehensive analysis that aims to scale to arbitrary binaries compiled from different languages, using different standard C libraries, etc.

\sys uses a heuristic in order to identify system call wrappers in the machine code (step \BC{G} in Figure~\ref{fig:overview}).
This heuristic works as follows.
Recall that Capstone/Angr is able to infer function boundaries from the disassembled machine code.
The key idea behind the heuristic is that we want to know whether the system call number is necessarily determined between the start a function containing a system call site, and the site in question.
If it is, we conclude that the function is not a wrapper.
Otherwise, then this system call number is determined elsewhere, for example by an argument of the function, and we conclude that it is a wrapper.
Based on these considerations, in order to detect wrappers, \sys employs a two-phases analysis that aims to minimize reliance on computationally-expensive symbolic execution.

The first phase is a simple use-define chain analysis that is fast but may yield false positives.
Starting from the system call site, we search backwards for \texttt{mov} instructions, up to the beginning of the function.
Then, by analyzing those instructions, \sys finds out whether \texttt{\%rax} can be determined, potentially by analyzing recursively other registers (if we encounter memory operands, we consider it to be undetermined).
This method can yield false positives, so if this first phase is positive, the function \emph{may} be a wrapper, and we confirm/disprove that hypothesis with a second phase leveraging symbolic execution.
\sys launches symbolic execution between the beginning of the function and the system call site, and if at this point, \texttt{\%rax} is symbolic, the function is definitively qualified as a wrapper.
In such case, the symbolic search also records which of the wrapper function parameter holds the system call identifier: it is either a register or a stack slot according to the programming language function call ABI.
The first phase is fast, whereas the second phase is potentially time-consuming.
This is why \sys performs symbolic analysis only if the first phase is positive.

Once a wrapper is detected, the system call identification process (\BC{H} in Figure~\ref{fig:overview}) for that particular system call site is slightly different from that of non-wrapper call sites.
To find the system calls made by the program through a wrapper, we start a similar search process as described above, with the difference that 1) the target code address we direct the symbolic execution toward is not the system call site but the first instruction of the wrapper; and 2) we aim to determine the value present in the register/stack slot holding the wrapper's parameter corresponding to the system call identifier, and not \texttt{\%rax}.

\subsection{Step 3: Dynamic Executables and Shared Libraries Processing}

\paragraph{Decoupled Executable/Library Analysis.}
\label{sec:dynamic-binaries}

As the majority of executables are dynamically compiled, it is necessary for \sys to be able to handle them.
Most of the symbolic methods that we use for static binaries can be reused to analyze shared libraries.
One of our goals when handling dynamically compiled executables and their shared library dependencies is to factorize the analysis of shared libraries.
Indeed, many libraries are used by multiple programs (e.g. \texttt{libc.so}, \texttt{libpthread.so}, etc.).
In order to avoid repeating, for each program using a shared library, some costly operations involved in processing the library, \sys implements a two-phase approach.
The first and computationally-expensive phase is done only once per library.
It includes the disassembly/CFG recovery, and system call sites location/wrapper detection/type identification for each function exposed by the library, i.e. steps \BC{D} to \BC{H} in Figure~\ref{fig:overview}.
The second phase is done during the analysis of the main dynamically compiled executable (step 3 in Figure~\ref{fig:overview}).
During this phase, we load (recursively) every needed shared library, we compute (recursively) the
reachable functions for every library, complete the over-approximation of each call-graph by computing active address taken and resolving indirect jumps, and deduce
the system calls potentially invoked by every function of those shared libraries that may be called by the executable.
This phase is executed once for each library dependency of a dynamically executable.

\paragraph{Library Analysis and Shared Interface.}

The analysis of a shared library (steps \BC{D} to \BC{H} in Figure~\ref{fig:overview}) is very similar to that of an executable and consists in the following:
1) listing the functions exposed by the shared library;
2) constructing the call-graph between the different functions of the shared library, potentially by taking into account external symbols;
3) detecting \texttt{address taken} functions and indirect call sites;
4) collect system call sites, detecting wrappers and handling indirect calls and;
5) inferring system call numbers.

\sys then outputs a result which will be referred to as the library's \emph{shared interface} (\BC{K} in Figure~\ref{fig:overview}).
This interface corresponds to a set of metadata that will later be used to construct a simple and direct correspondence between the library's exported functions (e.g. \texttt{printf}, \texttt{read}, etc. for the standard libc) and the system calls each of these functions can invoke.

Concretely, the shared interface corresponds to a json file.
It contains in particular
1) a function-level CFG of the library;
2) a list of system calls that can be invoked by each function;
3) the list of functions defining addresses taken and the addresses taken in question;
4) the list of functions corresponding to system call wrappers;
and 5) for each function calling a function exposed by another shared libraries, the list of such calls.

\paragraph{Dynamically Compiled Executable Analysis.}

The shared interface produced once per library is used during the analysis of the main executable binary or the analyses of other shared libraries.
While processing the main binary, when an external symbol is encountered (\BC{J} in Figure~\ref{fig:overview}), the system calls it can trigger are determined using the shared interfaces.

During this phase, \sys loads recursively every shared library needed by the executable, computes recursively the reachable exposed functions and active addresses taken of each library, refines the over-approximation of each call-graph with these addresses taken, and deduces the system calls that can be invoked by every exposed function of those shared libraries that may be called by the executable.
The first step is to determine an ordering of the nodes compatible with this DAG, which we implemented using a priority queue.
Then, following this order, \sys computes the reachable functions and active addresses taken of every shared library, by propagating the information of the reachable functions of its predecessors.
Next, following the reverse order, by resolving indirect calls with the identified active addresses taken, \sys deduces the complete call-graph of every shared library, and constructs a simple interface associating to every function exposed by the library a set of system calls that may be invoked when the function in question is called.
Then, this interface is used to construct recursively the interfaces of the other libraries.
Finally, we are able to analyze the executable, and to handle every external symbol in it.

Shared objects loaded at runtime through primitives such as \texttt{dlopen} (e.g. modules) are analyzed by \sys similarly to shared libraries.
We make the assumption that the user is able to identify the modules that can be loaded by an application's main binary: due to the nature of runtime loading mechanisms, it is very hard to identify them automatically.
It is also worth noting that the use of these mechanisms is relatively rare, e.g. in the set of 557 programs from the Debian repositories we consider in evaluation, less than 10\% use \texttt{dlopen}.

\subsection{Soundness and Performance Barriers}
\label{sec:soundness}

As mentioned earlier, similarly to all other works on binary-only system call identification~\cite{SYSFILTER, CHESTNUT}, \sys cannot totally guarantee soundness, i.e. the absence of false negatives.
This is due in part to the use of Angr which, in line with similar reverse-engineering tools, does not fully guarantee the correctness of the recovered CFG.
The complete and sound disassembly of arbitrary programs is an undecidable problem~\cite{CFG_UNDECIDABLE}, and we consider it to be orthogonal to our effort.
Further, please note that we successfully validate in evaluation (Section~\ref{sec:validation}) the correctness of filtering rules derived from \sys's analysis over a set of popular applications, with high-coverage workloads (full test suites).
Another barrier to soundness in \sys is our system call wrapper detection method, which is a heuristic in nature.
Note that it has been validated over wrappers implemented in different libc types (Glibc, Musl, Bionic) and languages (Go, Rust), over different versions.
Another issue related to performance is the fact that the symbolic execution search may never finish or run out of memory.
We demonstrate in evaluation that the success rate (no timeout/crash) of \sys on a large set of 500+ binaries is much higher than that of competitors.

\subsection{Phase Detection}
\label{sec:phases}

Observing that the set of system calls that can be invoked by a program may vary with its phases of execution, recent works~\cite{TEMPORAL} have presented the concept of temporal system call specialization, in which different filtering rules are applied to a process, one per phase of execution, for enhanced security.

Existing work uses dynamic analysis to detect phases and the different system calls that may be invoked during them.
Given the previously-mentioned downsides of dynamic analysis, with \sys we explore how efficient binary-level static analysis is at detecting such phases.
An intuitive method would be to navigate the CFG and merge into phases sets of nodes that contain system call sites and that are highly connected.
In practice, we observe that navigating the CFG to perform that operation is a very costly operation that does not scale well to medium/large programs.
Hence, we propose below an optimized method based on automata theory.

The key idea of this method is to transform the CFG and list of possible system calls issued at every system call site into a Deterministic Finite Automaton (DFA) in which states are sets of program's instruction (program's states) and transitions are the types of system calls used to switch states: the input alphabet of the automaton is the list of all system calls that can be invoked by the program.

\sys first creates a Non-deterministic Finite Automaton (NFA), by decorating the outgoing edges of every node containing a system call site with the list of system calls that can be invoked at that site.
All other edges are transformed into empty ($\epsilon$-) transitions.
At that stage the number of $\epsilon$-transitions in the NFA is very high as in the initial CFG only a small portion of nodes include system call sites.
The next step is to transform the NFA into an equivalent DFA that will be used for phase identification.
To that aim \sys uses a standard powerset construction algorithm.
That process eliminates $\epsilon$-transitions and ensures that every state has at most one outgoing transition of a given system call type.

\begin{figure}
    \center
    \includegraphics[width=0.45\textwidth]{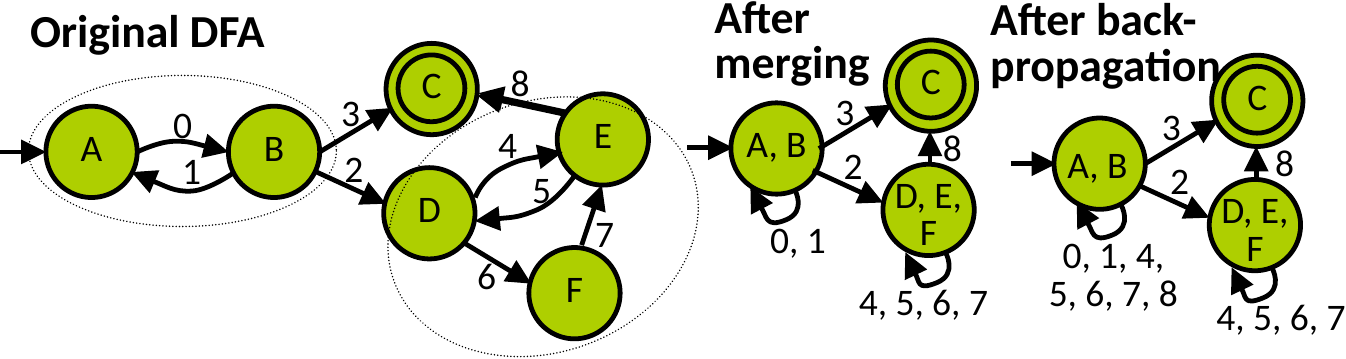}
    \caption{Automaton-based phase detection: states represent sets of basic blocks and transitions are system call invocations. Highly-connected states are merged to form phases, each with a system call allowed list (transitions from the phase).}

    \label{fig:automata}
\end{figure}

\sys then obtains a DFA in which highly-connected states can be merged to obtain phases as depicted in Figure~\ref{fig:automata}.
Each phase is characterized by the set of basic blocks composing it and a list of allowed system calls during that phase (transitions from that phase).
Outgoing (non-self) transitions represent system calls transitioning to other phases.
The final step consists in back-propagating authorized system calls to predecessor states.
Note that it is needed only when Linux's seccomp is used as the runtime filter.
Note that the example in Figure~\ref{fig:automata} is overly simplistic for the sake of illustration.
In reality, the number of states is higher, and they are much more connected due to the CFG overestimation and to the NFA to DFA transformation process.
This transformation process is highly optimized, and the automaton-based method is much faster than the graph navigation one:
for a simple hello world static binary (32K basic blocks of machine code), it takes 41 seconds vs. 700 seconds for the intuitive CFG navigation technique presented earlier.
For a medium-sized program (Nginx, 83K basic blocks), that difference is 20 minutes vs. 4 hours.

Existing works regarding phase-based filtering~\cite{TEMPORAL} assume access to the application sources so that an expert can manually place in the code calls to install new rules at phase boundaries.
Given \sys's assumptions (no source access), enforcing phase-based filtering would have to be achieved differently.
The most efficient way would be to slightly modify the filtering mechanism in the kernel~\cite{STATEFUL_SECCOMP} (e.g. seccomp) to detect phase changes by monitoring, when a system call is invoked, the system call type and the application's return address.
Another possibility would be a user-space monitor~\cite{BASTION}.
These solutions would also allow overcoming seccomp's limitation of being able to subsequently install only stricter rules, hence to drop the back-propagation step for more flexible and secure phase-based policies.
Still, we believe modifying seccomp/developing a user-space monitor falls out of the scope of this paper that focuses on the problem of system call identification.

\section{Evaluation}
\label{sec:evaluation}

In this evaluation we answer the following questions:
\begin{compactitem}

\item Is \sys valid, does it suffer from false negatives?
    (\S \ref{sec:validation})

\item How precise is \sys, i.e. to what extent does it suffer from false positives?
    How does this overestimation compare to state-of-the-art competitors?
    (\S \ref{sec:precision})

\item What are the execution times and memory footprints of \sys's analysis?
    (\S \ref{sec:exec-time-rss})

\item How strict would be the filtering policies that can be derived from \sys's phase detection analysis?
    (\S \ref{sec:phases-eval})

\item How efficient would be the filtering rules derived from \sys's analysis at protecting a large set of programs against a set of real-world CVEs?
    (\S \ref{sec:secu-eval})

\end{compactitem}

In several experiments we compare \sys to Chestnut~\cite{CHESTNUT} and SysFilter~\cite{SYSFILTER}.
These competitors are the most relevant because, similarly to \sys and contrary to other works~\cite{ABHAYA, SAPHIRE, BASTION}, they do not assume access to application/library sources.

\subsection{Validation}
\label{sec:validation}

Existing works validate the correctness of the filtering rules derived from their analysis by relying on a form of dynamic analysis, as other approaches either lack precision (source-level analysis) which defeats the purpose, or are simply not doable (manual analysis).
Some competitors check if standard benchmarks~\cite{CONFINE} or test suites~\cite{SYSFILTER} succeed while enforcing the filtering rules.
This does not allow to reason about false positives (system calls identified by the tools but not actually issued by the application), hence we take a different approach.
We compare for a set of applications the system calls identified by \sys to a ground truth, in order to detect false positives as well as possible false negatives (system calls present in the ground truth but not detected by the analysis).
Similar to existing works~\cite{CONFINE, SYSFILTER, CHESTNUT}, the ground truth is established using dynamic analysis: we trace with \texttt{strace} at runtime the system calls invoked by each considered application when running its entire test suite.
We select 6 applications: Redis, Nginx, HAProxy, Memcached, Lighttpd and SQLite.
Similarly to prior works, we assume that their test suites achieve sufficient coverage for dynamic analysis to produce a ground truth of good quality~\cite{CHESTNUT}.
These applications are highly popular and widely deployed in production, hence they need to be extensively tested with high-coverage test suites.
Note that some applications (e.g. Nginx) load modules at runtime through \texttt{dlopen}.
These are processed alongside the main binary, similarly to shared libraries.

\begin{figure}
	\center
	\includegraphics[width=0.45\textwidth]{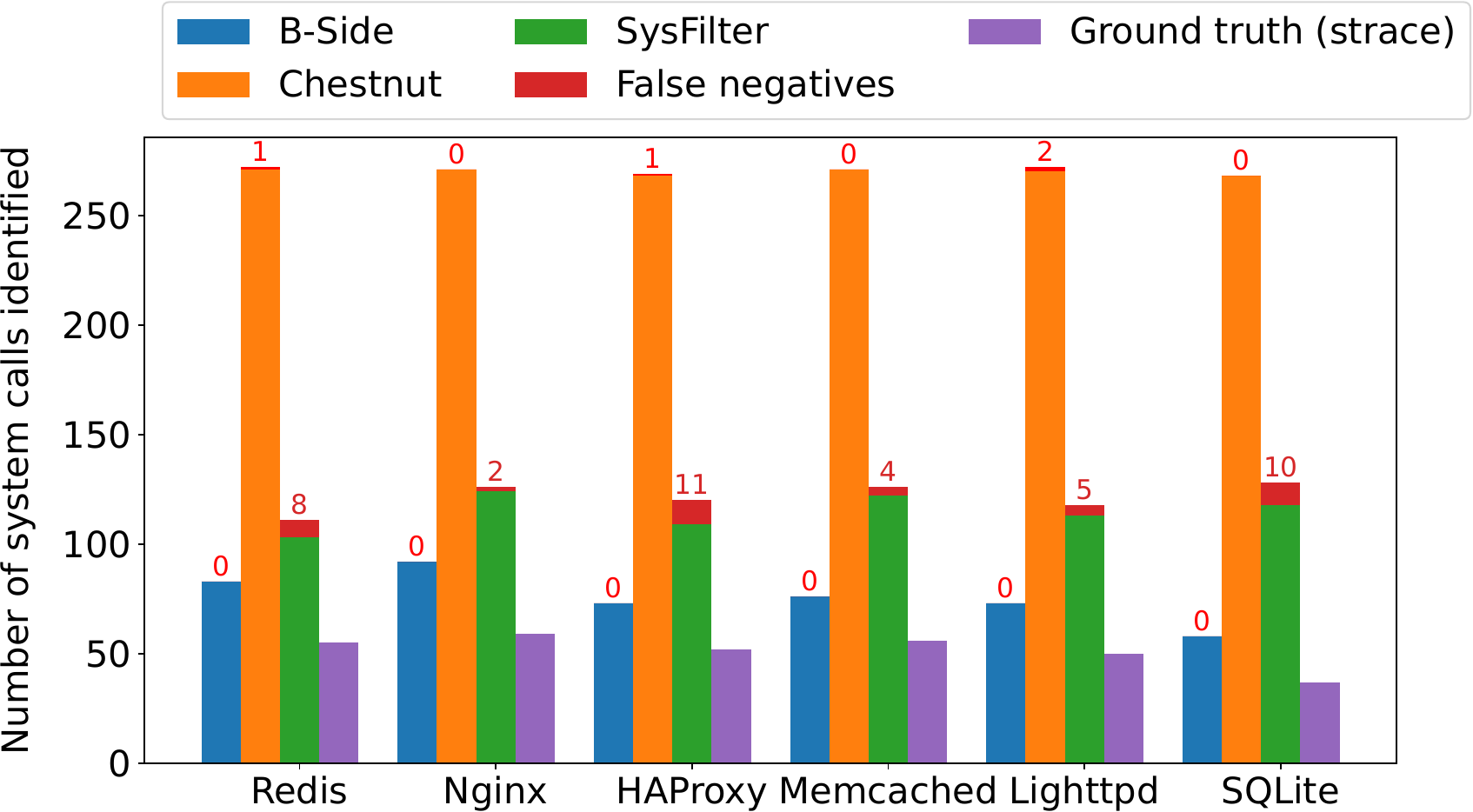}
\caption{System calls identified by \sys, Chestnut, SysFilter, and \texttt{strace} (on test suites) on 6 applications.}

	\label{fig:validation}
\end{figure}

The results are on Figure~\ref{fig:validation}.
We observe that \sys does not present any false negatives, while the competitors do.
We believe that this is mainly due to the lack of system call wrapper handling in both Chestnut and SysFilter, which leads to system calls that will be missed by the analyses.
Chestnut presents a lower amount of false negatives compared to SysFilter, which is mainly due to the very high number of system call identified (more than 250 for each application), denoting a high amount of false positives.
We also successfully validated \sys on these applications with an alternate C library, Musl, both in statically and dynamically compiled modes.

\subsection{Precision}
\label{sec:precision}

\paragraph{Popular Applications.}

\begin{table}
	\centering
    \caption{$F_1$ scores for \sys, Chestnut and SysFilter over the 6 binaries considered for validation.}
    \includegraphics[width=.46\textwidth]{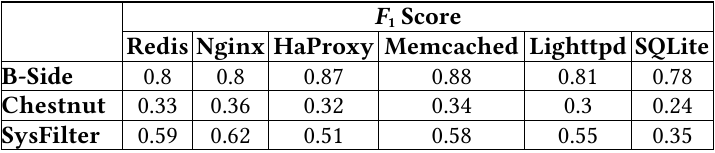}
	\label{tab:precision}
\end{table}

Table~\ref{tab:precision}, presents the $F_1$ score (harmonic mean of precision and recall) for \sys, Chestnut and SysFilter on each of the binaries considered in the validation.
\sys's score is consistently higher than the competitors': its average $F_1$ score over all applications is 0.81, vs. 0.31 for Chestnut and 0.53 for SysFilter.
The main reason for that is the high number of false positives shown by these competitors, as illustrated on Figure~\ref{fig:validation}.
In terms of dangerous system calls~\cite{CHESTNUT}, we confirmed that \sys is able to filter out \texttt{execve} on Nginx/Memcached, and \texttt{execveat} on all popular applications.

\paragraph{Debian Binaries.}

To measure precision at scale, we execute it over a vast set of programs: we selected from the Debian 10 x86-64 repositories (\texttt{main}, \texttt{contrib} and \texttt{non-free}) every ELF executable containing at least one occurrence of the \texttt{syscall} instruction.
For dynamically-compiled binaries, we also select the corresponding shared library dependencies that include this instruction.
We gather a total of 557 executables: 231 static binaries and 326 dynamically-compiled with their 59 shared library dependencies.
We confirmed by exploring the corresponding source packages that these binaries come from the compilation of many languages: C, C++, Haskell, Go, etc.
\sys is compared to Chestnut and SysFilter over all binaries.
We compare the number of executables for which each system succeeds/fails, as well as the number of system calls identified.

\begin{table}
	\centering
    \caption{\sys vs. Chestnut and SysFilter on 557 binaries extracted from the Debian 10 repositories. For each system considered we sum up the number of static and dynamic binaries upon which the analysis succeeds to aggregate the number of successes over the total set of binaries considered, and proceeded similarly for the failures.}
    \includegraphics[width=.47\textwidth]{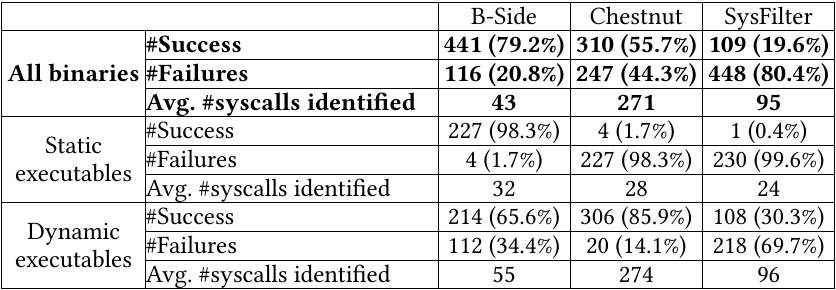}
	\label{tab:debian-res}
\end{table}

The results are summarized in Table~\ref{tab:debian-res}.
On dynamic binaries, \sys's precision is superior: it identifies on average 55 system calls vs. 274 for Chestnut and 96 for SysFilter.
Figure~\ref{fig:validation} details the exact number of system calls identified by each solution for each of the 6 applications used in validation.
Chestnut always identifies more than 268 system calls, and SysFilter more than 103.
\sys reports between 49 and 84 (58 and 92 when accounting for false negatives).
These numbers are close to the ground truth established from dynamic analysis on test suites, highlighting the low number of false positives in \sys's analysis.
Chestnut and SysFilter are not as precise and show many false positives.

\begin{figure}
	\center
	\includegraphics[width=0.45\textwidth]{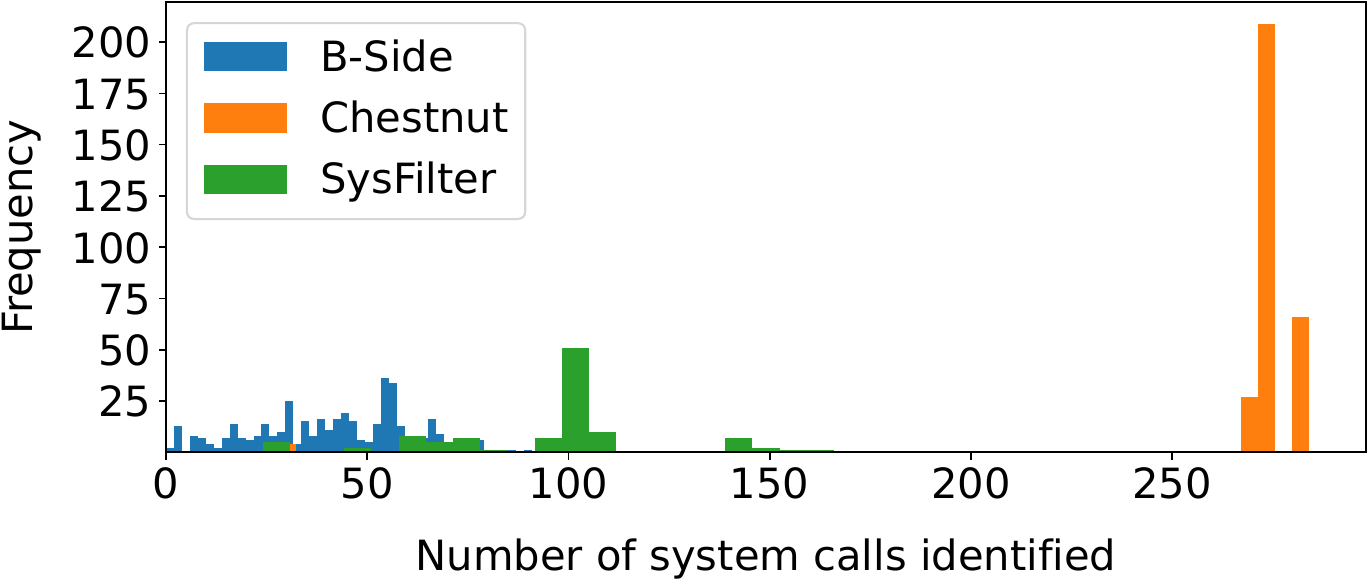}
    \caption{Distribution histogram of the number of system calls identified by \sys, Chestnut, and SysFilter over the set of Debian binaries upon which each tool successfully runs.}

	\label{fig:distribution}
\end{figure}

\begin{figure*}
    \includegraphics[width=\textwidth]{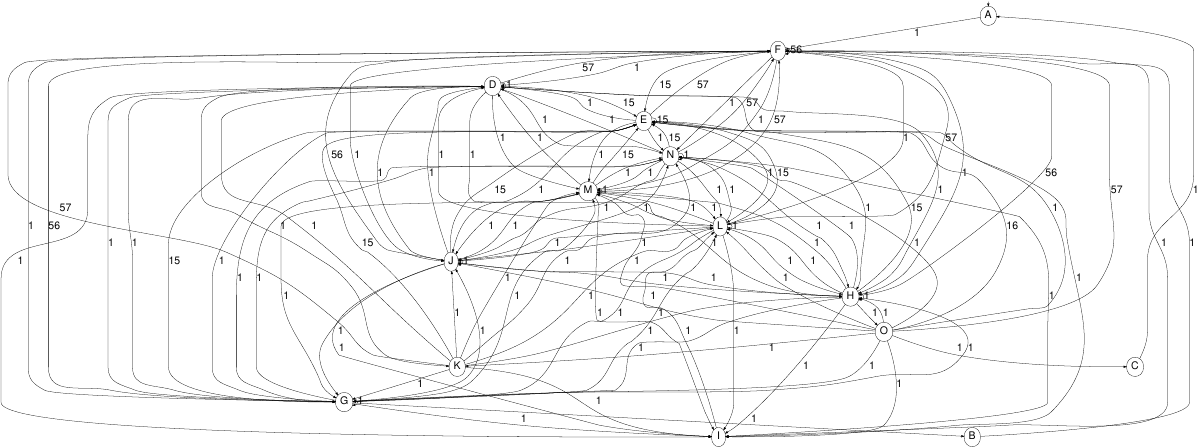}
    \captionof{figure}{Automaton representing Nginx phases of execution extracted by \sys.
        Note that for clarity reasons, for each state we do not have one edge per allowed system call, but rather one edge per possible destination phase, labeled with the number of different system calls types triggering that transition.}

    \label{fig:nginx-phases-automaton}
\end{figure*}

Figure~\ref{fig:distribution} presents the distribution of the number of system calls identified for all applications in our set.
For the vast majority of executables Chestnut identifies around 270 system calls, and there are very few variations of that number with applications.
SysFilter also reports around 100 system calls in most cases.
On the contrary, for most applications \sys generally identifies considerably fewer system calls (between 1 and 90), and the numbers vary highly on a per-application basis.
This further shows the higher precision of \sys and its ability to identify a close superset of the system calls that may be invoked by an application at runtime.

Although Chestnut and SysFilter identify fewer system calls than \sys in static binaries (respectively 28 and 24 vs. 32), they fail on most static binaries (respectively 227/231 and 230/231), making the comparison relatively meaningless.
We investigated the reasons behind Chestnut's failure on static binaries and found that it is linked to its lack of management of system call wrappers.
Concerning SysFilter, its failure is due to its lack of support for non-PIC binaries.

\paragraph{Failures.}
We observe that \sys fails on part (112/326) of dynamically compiled executables, due to the analysis not finishing within the maximum time window we defined (3 hours max for each program).
Studying these timeouts, the vast majority of them (73\%) happen during the CFG construction phase which is generally the longest one (see Table~\ref{tab:exec-time-rss}).
For this phase we entirely rely on Angr's CFG recovery feature.
As such we consider that issue orthogonal to \sys and hope Angr will be enhanced to support faster CFG recovery for most binaries.
A few timeouts (15\%) happen during the core system call identification phase leveraging symbolic execution.
The amount of paths symbolically explored when the analysis times out during the system call identification phase varies according to the binary, but is generally high (thousands of basic blocks explored).
Giving more time to the analysis to complete would address the problem, although it is hard to determine how much time would be needed.
A small amount of timeouts (12\%) also happen during the execution of our wrapper identification analysis.
Optimizing this aspect of \sys will be a priority in future works.
As one can observe on Table~\ref{tab:exec-time-rss}, the wrappers and system calls identification phases are generally shorter than the CFG recovery, which explain the lower amount of timeouts during wrapper/system call identification.

\subsection{Execution Time and Memory Footprint}
\label{sec:exec-time-rss}

\begin{table}
	\centering
    \caption{\sys's analysis execution time, memory footprint (peak resident set size), and number of basic blocks executed symbolically during the system call identification phase on the applications used in validation.}
    \includegraphics[width=.47\textwidth]{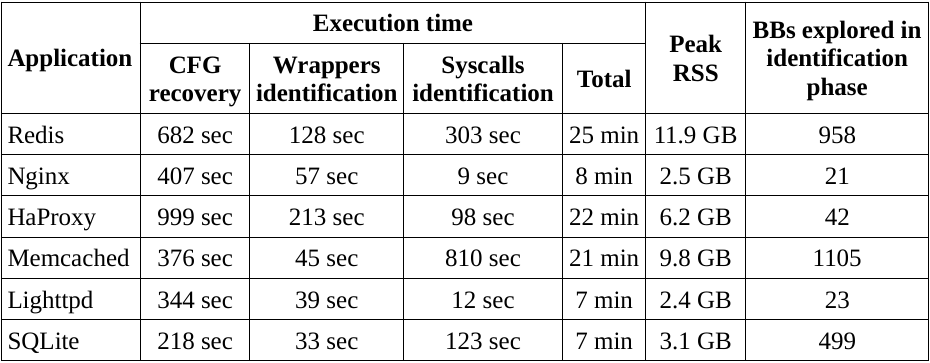}
	\label{tab:exec-time-rss}
\end{table}

Table~\ref{tab:exec-time-rss} presents the execution time and peak resident set size of \sys's analysis when applied to each of the applications used in our validation (\S \ref{sec:validation}).
Execution times are in the order of minutes, ranging from 6.5 minutes (SQLite) to 26 minutes (Redis).
The table also presents the execution time of the three main steps of the analysis: as one can observe, the time taken by Angr to recover the CFG generally dominates.
Note that the sum of the 3 main phases of the analysis (CFG recovery, wrappers identification, and system call identification) is inferior to the total analysis time as there are other steps involved (e.g. loading the binary) which execution time is not presented here.
The memory footprint varies from 2.4 GB (Lighttpd) to 11.9 GB (Redis).
Execution time and memory footprint are a one time cost to generate a filter for a given binary with \sys, and as such optimizing these was not a primary objective for our work.

\subsection{Phase Detection}
\label{sec:phases-eval}

To evaluate the effectiveness of \sys's phase detection technique, we ran it on the 6 applications presented in the validation.
Here we describe the results for Nginx, but note that the observations are similar for all 6 applications.

Figure~\ref{fig:nginx-phases-automaton} presents a simplified view of the automaton extracted by \sys on Nginx (see Section~\ref{sec:phases-eval}) before the back-propagation step (see \Cref{sec:phases}) and representing its various phases of execution.
In this figure, each state corresponds to a phase.
For the sake of clarity, each edge outgoing from a phase $P$ to another phase $P'$ represents a set of system calls allowed in $P$, which invocation triggers a transition to $P'$.
The edge's label indicates the number of system calls composing the set in question.
Note that $P$ and $P'$ can correspond to the same phase.
The number of system calls allowed in each phase corresponds to the sum of the labels of each outgoing edge.
As a reminder the total number of system calls identified in Nginx' binary by \sys is 93.

\begin{table}
    \center
    \caption{Nginx phase automaton: each cell gives the size of a set of system calls allowed in the source phase and triggering a transition to the destination phase.
    The last columns give the total number of system calls allowed in each phase and the total size of the basic blocks composing the phase.}
    \includegraphics[width=.47\textwidth]{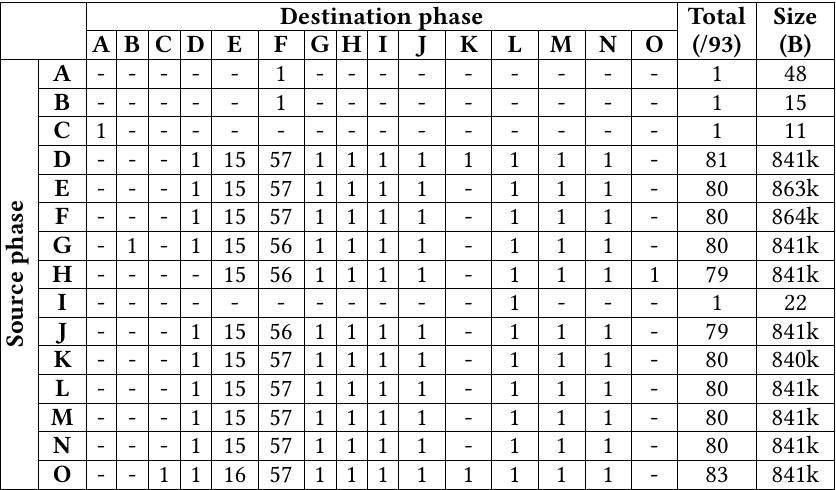}
    \label{tab:nginx-phases}
\end{table}

To further understand the output of our phase detection analysis, we present in \Cref{tab:nginx-phases} a summary of the automaton: the number of detected phases and, to determine how strict each phase is, the number of identified system calls per phase.
\sys identifies 15 phases for Nginx.
In this table cells give the number of system calls triggering a transition between a source (row) and destination (column) phases, each phase being a state of the automaton.
The total column gives the number of system calls allowed in each phase, i.e. its strictness.
To estimate how much time the program would spend in each phase, we also report in the last column the code size of each phase by summing the sizes of all basic blocks composing each phase.
Note that due to the method we use to make the automaton deterministic, a basic block can belong to several phases.

We can categorize phases into two classes.
Small phases (A-C and I) are very strict with a single system call allowed, but also small (a few bytes of code), hence it is likely that the program will spend the majority of its execution outside such phases.
Large phases (D-H and J-O) allow a higher number of system calls: 79 to 83, representing respectively 85\% and 89\% of the total amount of system calls identified in the binary (93).
Such phases have a much higher size (840-860 KB) and the program will spend most of its execution in them.
Hence, for Nginx, using a phase-based filtering system would lead to an average increase in strictness of 11 to 15\%.
We observed similar numbers for the other 6 applications studied.

\subsection{Security Support}
\label{sec:secu-eval}

We aim to understand how efficient the filtering rules that can be derived from \sys's analysis would be at protecting against a set of real-world CVEs.
We extracted from the literature~\cite{CONFINE, SYSFILTER, KITE} a list of CVEs involving various (Linux) kernel-level vulnerabilities that can be triggered through system calls.
Each CVE is mapped to the system call(s) triggering it.
For a CVE $C$ triggered by a system call $S$, if in a program $P$ \sys does not identify $S$, then one could derive a filtering policy for $P$ precluding $S$ and thus protecting against $C$.
We consider the list of 557 Debian binaries from Section~\ref{sec:precision}, and compute the percentage of programs from that set that would be protected against each CVE given a filtering rule derived from \sys's analysis.

\begin{table}
    \caption{Percentage of Debian binaries protected by filtering rules derived from \sys's analysis for a set of CVEs.}
    \center
    \includegraphics[width=.47\textwidth]{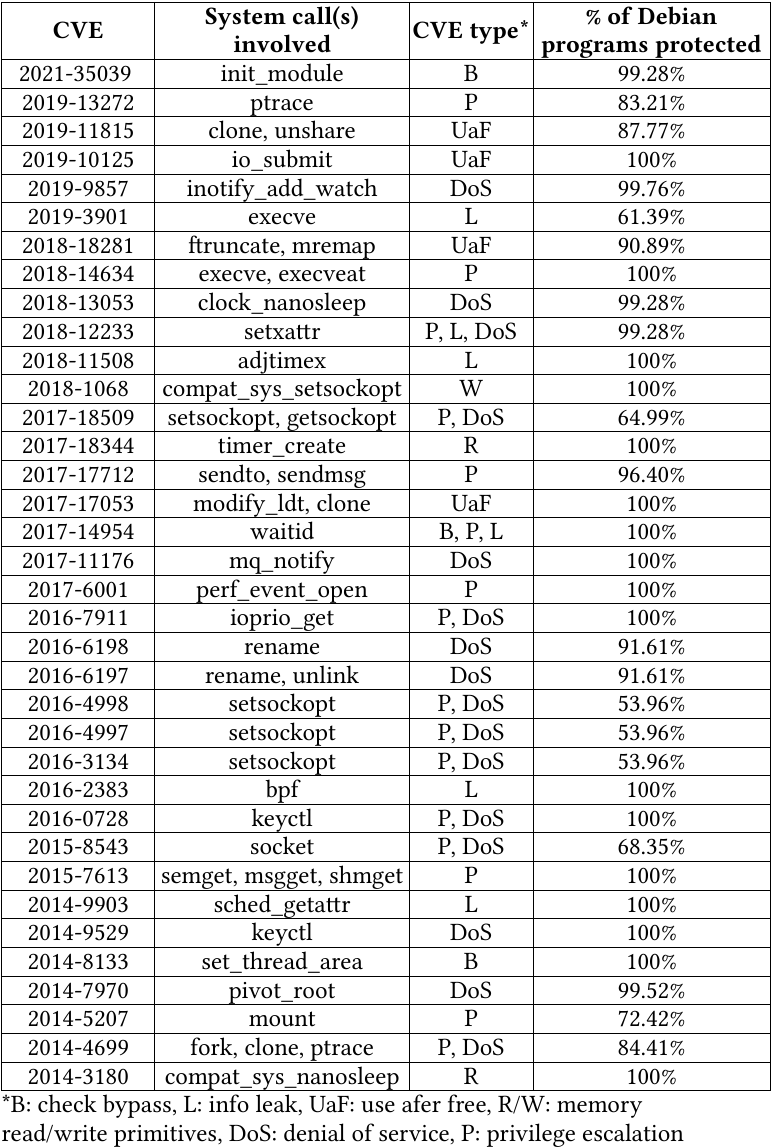}
    \label{tab:secu-eval}
\end{table}

We gathered a list of 36 CVEs for which system calls are involved in the attack process.
These CVEs are extracted from the following papers: SysFilter~\cite{SYSFILTER}, Confine~\cite{CONFINE}, as well as Kite~\cite{KITE}, and filtered out CVEs prior to 2014 for space reasons.
They are presented in Table~\ref{tab:secu-eval}, along with a description of their impact, and the percentage programs protected.
We find that on average over all CVEs, a filtering rule derived from \sys's analysis would protect 90.33\% of the Debian binaries we consider.
For 16 CVEs, 100\% of the binaries are protected, and there is no CVE for which less than 54\% of the programs are protected.
The percentage of binaries protected depends on the popularity of the system call triggering the vulnerability.
For CVEs caused by rarely-used system calls, e.g. CVE-2019-10125 relying on \texttt{io\_submit} and CVE-2016-2383 relying on \texttt{bpf}, a filtering rule will protect the majority of binaries (100\% for these two).
With popular system calls, such as \texttt{setsockopt} from CVE-2016-4998, fewer applications are protected (54\%).
Still, overall, these numbers show that a high number of vulnerabilities can be avoided through filtering rules derived from \sys's analysis.

\section{Conclusion}\label{sec:conclusion}
System call filtering brings the need for automated binary-only system call identification techniques.
\sys is a static analysis tool identifying a superset of the system calls an executable may invoke at runtime, without assuming access to sources.
It leverages symbolic execution as well as a heuristic to detect system call wrappers to provide precise results.
It can also detect phases of execution in a program in which different filtering rules can be applied.
\sys offers a high precision: validated over six popular applications, \sys's average $F_1$ score is 0.81, vs. 0.31 and 0.53 for competitors.
Over more than two hundred dynamically-compiled binaries, \sys identifies an average of 55 system calls, vs. 274 and 96 for competitors.

\section*{Acknowledgments}
We thank the anonymous reviewers and our shepherd, Emanuel Onica, for their insightful feedback and invaluable help increasing the paper's quality.
This work was partly funded by Orange Innovation; the UK's EPSRC grants EP/V012134/1 (UniFaaS), EP/V000225/1 (SCorCH), EP/X015610/1 (FlexCap); and the UK Industrial Strategy Challenge Fund (ISCF) under the Digital Security by Design (DSbD) Programme delivered by UKRI as part of the projects 10017512 (MoatE) and 75243 (Soteria).

\bibliographystyle{ACM-Reference-Format}
\bibliography{bib}

\end{document}